\documentstyle[epsfig,eqsecnum,prl,aps]{revtex}

\newcommand{\doublespace}{
  \renewcommand{\baselinestretch}{1.75}
  \large\normalsize}
\begin{document}
\doublespace
\large

\vspace{.2in}

\title {Zero-Bias States and the
Mechanism of the Surface $d\rightarrow d+is$ Transition}

\author{\v Simon Kos}
\address{University of Illinois, Department of Physics\\ 1110 W. Green St.\\
Urbana, IL 61801 USA
\\E-mail: s-kos@uiuc.edu
}

\maketitle

\begin{abstract}

We study the physical mechanism of the surface
$d\rightarrow d+is $ transition proposed
as the interpretation of results of tunneling experiments into $ab$ planes
\cite{covington97}.
We base our argument on 
first-order perturbation theory and show that the zero-bias states drive the
transition. We support the argument by various estimates and consistency
checks.

\end{abstract}

\def\argsinh{\mbox{argsinh}}
\def\sgn{\mbox{sgn}}
\section{Introduction}

It has now been firmly established 
that the order parameter $\Delta $ in
homogeneous cuprate superconductors has a $d$-wave symmetry \cite{dale95}. 
It follows
that inhomogeneities can scatter quasiparticles between directions that
experience opposite signs of $\Delta $. This effect is strongest at a
specularly reflecting (110) surface, because $\Delta $ changes sign along
each quasiclassical trajectory upon reflection (see Fig. 1). As a consequence
of the Atiyah-Patodi-Singer index theorem
\cite{atiyah}, the Andreev equation along each
such trajectory has in its spectrum a bound state at exactly zero energy,
irrespective of the detailed shape of the potential
\cite{hu}. Collectively, these states then
make up a peak in the tunneling spectra, which has been observed 
experimentally \cite{geerk,lesueur,covington96}.
\vskip 5 mm
\begin{figure}
\epsfig{file=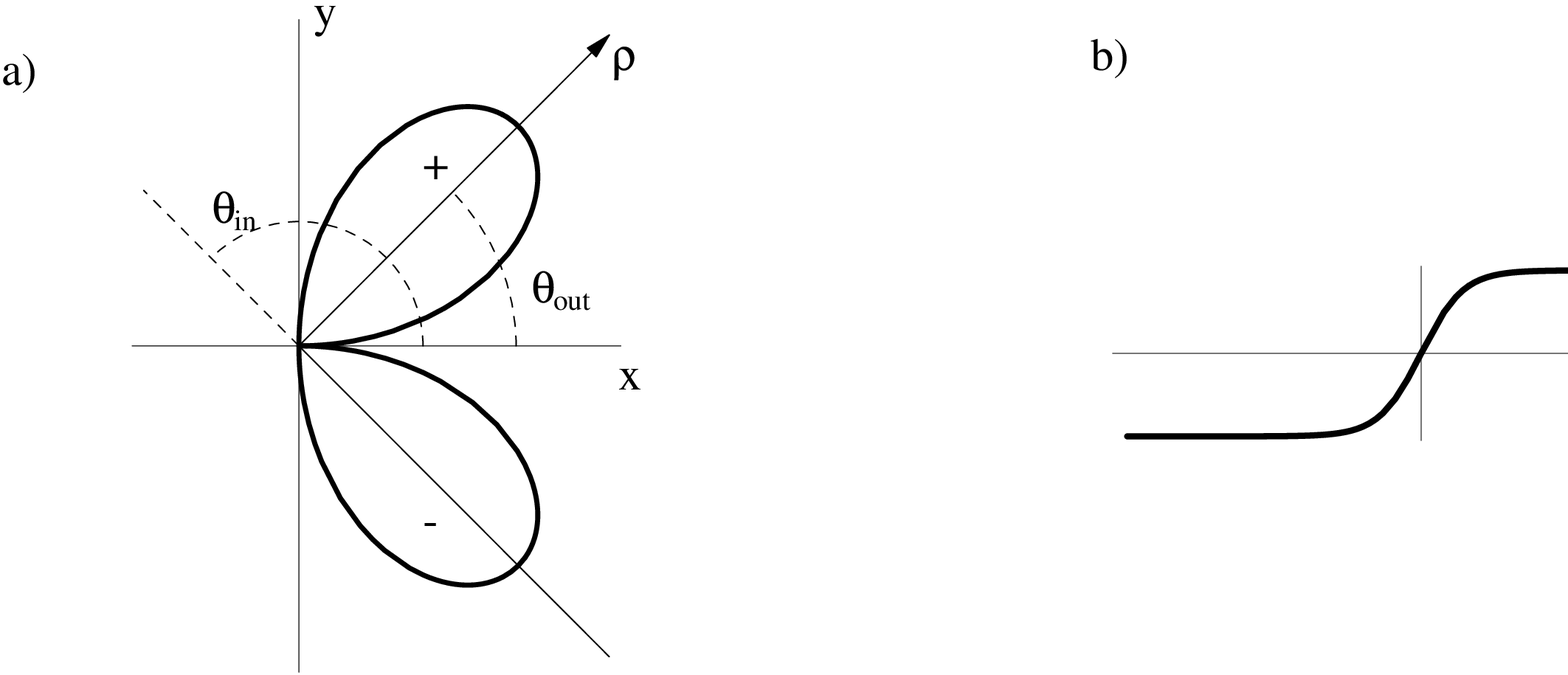,width=12.0cm}
\caption{a) A schematic picture of the normal metal--superconductor junction in
the (110) direction with a typical quasiclassical trajectory.
\newline
b) A schematic graph of the pairing potential along the trajectory in a).}
\end{figure}

It was predicted theoretically \cite{matsumoto}
and confirmed experimentally
\cite{covington97} 
that below a certain temperature of the order of 1K, this peak
of the zero-bias states (ZBSs) splits. The theoretical interpretation
of this splitting is that via a sub-dominant pairing interaction of a
different symmetry, say $s$, a subdominant order parameter is induced close
to the surface. It is phase-shifted by $\pi /2$ relative to the dominant
$d$-wave, which gives the total order parameter ``$d+is$'' symmetry and
indicates broken time-reversal symmetry.

The full self-consistent calculations in the Eilenberger formalism that are
based on this interpretation \cite{fogelstrom,rainer} 
are in a good quantitative agreement
with the experimental data. However, the mechanism of the transition is not
manifest in the numerical solutions of the Eilenberger equations. We believe
that understanding of the basic physics of the $d\rightarrow d+is$ transition is
especially needed now in light of recent experiments that call this 
interpretation into question \cite{dale99}. 
That is the purpose of the work presented
in this paper.

Below, we show by a simple argument based on first-order perturbation theory
that the degrees of freedom driving the $d\rightarrow d+is$ transition are
the ZBSs, and that we can neglect the effect of all the remaining states.
Hence, to understand the mechanism of the transition, we have to deal
only with the ZBSs, which is convenient 
since these states are least sensitive to the
unknown surface details.

The paper is organized as follows: In Section \ref{2}, 
we demonstrate our strategy
on the familiar case of BCS instability. The main argument is presented in
Section \ref{3} 
after we have extended the BCS formalism to inhomogeneous systems
and non-$s$-wave pairing. Based on this argument, we calculate $\Delta $ at
$T=0$ in Section \ref{4} and estimate the transition temperature to the $d+is$
state in Section \ref{5}. In Section \ref{6}, 
we discuss the surface current. Finally,
we discuss our results in Section \ref{7}.

\section{BCS Instability}
\label{2}

There are various ways to consider the energetic costs and benefits of the
transition to the superfluid state. The one that has proven useful in our
study of the $d\rightarrow d+is$ transition is to decouple the attractive
four-fermion interaction by the Hubbard-Stratonovich (HS) transformation, and
to make a saddle-point (mean-field) approximation. That way, we break up 
the total free energy of the system into free energy of  single particle
states, which is lowered by the gap $\Delta $, and the extra term from
the HS transformation, which grows (quadratically) with $\Delta $. We then see
that at small enough $T$, the system favors transition to the superfluid
state.

We will demonstrate this on the familiar BCS case. The model 
Hamiltonian is
\footnote{Since there is no universal convention as to whether the
attractive interaction term should have a plus sign with a negative coupling
constant $V$ or a minus sign with positive $V$, we use $|V|$ which is
unambiguously positive.}
\begin{equation}
H=\sum\limits _{{\bf k},\sigma } \epsilon _{\bf k}
c^{\dag }_{{\bf k},\sigma } c_{{\bf k},\sigma }-
|V| \sum\limits_{\bf k, k'}
c^{\dag }_{{\bf k} \uparrow } c^{\dag }_{{-\bf k} \downarrow }
c_{{-\bf k '} \downarrow} c_{{\bf k '} \uparrow },
\end{equation}
which gives rise to the partition function
\begin{equation}
Z=\int {\cal D} \overline{c}_{{\bf k} \sigma } {\cal D} c_{{\bf k} \sigma }
e^{-\int\limits_0^{\beta } d\tau \left[ 
\sum\limits_{{\bf k} \sigma } \overline{c}_{{\bf k} \sigma }
(\partial _{\tau } + \epsilon _{\bf k} ) c_{{\bf k} \sigma } -
|V| \sum\limits_{\bf k, k'}
\overline{c}_{{\bf k} \uparrow } \overline{c}_{{-\bf k} \downarrow }
c_{{-\bf k '} \downarrow} c_{{\bf k '} \uparrow }\right]},
\end{equation}
where the $c$'s are now $\tau -$dependent Grassmann numbers. We perform the
HS transformation by multiplying the partition function by the (infinite)
constant
$$
\int {\cal D} \overline{\phi }_{\bf k} {\cal D} \phi _{\bf k}
e^{-|V| \int\limits_0^{\beta } d\tau \sum\limits_{\bf k, k'}
(\overline{\phi }_{\bf k} - 
\overline{c}_{{\bf k} \uparrow } \overline{c}_{{-\bf k} \downarrow })
(\phi _{\bf k'} - c_{{-\bf k '} \downarrow} c_{{\bf k '} \uparrow })},
$$
so
\begin{equation}
Z=\int {\cal D} \overline{\phi }_{\bf k} {\cal D} \phi _{\bf k}
{\cal D} \overline{c}_{{\bf k} \sigma } {\cal D} c_{{\bf k} \sigma }
e^{-S},
\end{equation}
where
\begin{equation}
\label{action}
S=\int\limits_0^{\beta } d\tau \left[ \sum\limits _{{\bf k}}
\pmatrix{\overline{c}_{{\bf k} \uparrow } & c_{{-\bf k } \downarrow}}
\pmatrix{ \partial _{\tau } + \epsilon _{\bf k} & \Delta \cr
\overline{ \Delta } & \partial _{\tau } - \epsilon _{\bf k}}
\pmatrix{c_{{\bf k} \uparrow } \cr \overline{c}_{{-\bf k} \downarrow }}
+ {|\Delta |^2 \over |V|} \right],
\end{equation}
where we defined
$$
\Delta = -|V| \sum\limits_{\bf k} \phi _{\bf k}.
$$
From the action (\ref{action}), we can read off that in the mean-field
approximation, the total free energy of the system is
\begin{equation}
\label{freeenergy}
F(|\Delta |)= \sum\limits_{\bf k} ({\cal F}(E_{\bf k}) +
{\cal F}(-E_{\bf k}))+{|\Delta |^2 \over |V|}
\end{equation}
upon minimization with respect to $|\Delta |$. Here,
\begin{equation}
\label{F1}
{\cal F}(E) = -T \ln (1+e^{-E/T}),
\end{equation}
and 
$$
E_{\bf k} = \sqrt{ \epsilon _{\bf k}^2 + |\Delta |^2}.
$$

We see the instability most clearly at $T=0$, where 
$F(|\Delta |)=E(|\Delta |)$. Then
$$
{\cal F}(E) = \theta (-E) E,
$$
so
\begin{equation}
E(|\Delta |)- E(0) = N(0) \int\limits_{\omega _D}^0
d\epsilon (-\sqrt{\epsilon ^2 + |\Delta |^2} -\epsilon )
+ {|\Delta |^2 \over |V|},
\end{equation}
where $N(0)$ is the density of states at the Fermi level, and
$\omega _D$ is the Debye frequency. Direct calculation shows that
the integral behaves as $|\Delta |^2 \ln {|\Delta | \over \omega _D}$
for $|\Delta |\rightarrow 0$, whose non-analytic decrease will win
over the analytic increase of the second term for small enough $\Delta $,
no matter how weak the attractive interaction $|V|$ is. By the same
calculation, we can also see that the integral becomes analytic if we
do not integrate $\epsilon $ all the way up to zero, but to a finite negative
energy. This means that the states close to the Fermi energy drive
the BCS transition---they benefit most from opening of the gap $|\Delta|$.
Similarly, we shall see that the states at zero energy, that is the
ZBSs, will drive the $d\rightarrow d+is$ transition.

So far, the argument has shown the BCS instability only at $T=0$.
At finite temperatures,
$$
{\cal F}(E_{\bf k}) + {\cal F}(-E_{\bf k}) =
-T\ln (2+2\cosh {E_{\bf k} \over T}),
$$
which, upon expansion in powers of $|\Delta |^2$, gives
\begin{equation}
F(|\Delta |)- F(0) = |\Delta |^2
\left({1\over |V|}- N(0) \int\limits_{-\omega _D}^0
{d\epsilon \over \epsilon} \tanh {\epsilon \over 2T} \right)
+ O(|\Delta |^4).
\end{equation}
This shows that the system is unstable to the BCS transition at 
temperatures below $T_c$ that satisfies
\begin{equation}
{1\over |V|}- N(0) \int\limits_{-\omega _D}^0
{d\epsilon \over \epsilon} \tanh {\epsilon \over 2T_c}=0.
\end{equation}
In a similar way, we shall see below that $T_{s}$, the transition
temperature into the $d+is$ state, is finite.

\section{The $d+is$ Instability}
\label{3}
\subsection{Formalism}

We now need to develop the formalism that will enable us to extend the
strategy from Section \ref{2} 
to the $d+is$ case. We shall consider a single (2-dimensional) CuO plane, and
model it by the Hamiltonian
\begin{equation}
\label{fullham}
H  =  \int d^2 r \sum\limits_{\sigma = \uparrow , \downarrow}
\psi ^{\dag} _{\sigma } ({\bf r})
\epsilon (-i \nabla ) \psi _{\sigma }({\bf r})
 +  \int d^2 r d^2 r' V({\bf r- r'})
\psi _{\uparrow } ^{\dag } ({\bf r})
\psi _{\downarrow} ^{\dag} ({\bf r'})
\psi _{\downarrow }({\bf r'})
\psi _{\uparrow }({\bf r}),
\end{equation}
where $\epsilon $ is the band energy and $V$ is the short-range interaction
responsible for pairing. What makes this difficult problem tractable is the
separation of energy scales (the Fermi energy $E_F$ is much bigger than the
superconducting gap $\Delta $), which gives rise to separation of length scales
$\lambda _F$ (Fermi wave length) and $\xi $ (the coherence length). We may,
therefore, expand in powers of the small parameter $\lambda _F/\xi $; keeping
the lowest non-trivial order is called the quasiclassical approximation. This
procedure is usually done at the level of Green's function 
\cite{Eilenberger,ReinerSerene}, which are thus
transformed into Eilenberger functions that satisfy transport-like equations.

Since we want to understand the $d\rightarrow d+is$ transition in terms of
quasiparticle eigenstates rather than Green's functions, we will perform this
separation of scales at the operator level instead. We denote as $2\Lambda $
the width of the shell around the Fermi surface containing the states that
take part in the pairing (see Fig. 2). We then factor out the fast 
Fermi-surface oscillations and define the slowly varying field operator
$\psi _{\sigma , \theta } ({\bf r})$ \cite{stone} by
\begin{eqnarray}
\label{qcfield}
\psi _{\sigma } ({\bf r })& = &
\int {d^2k\over (2\pi )^2} c_{{\bf k}\sigma }
e^{i {\bf k} \cdot {\bf r}} \nonumber \\
& \simeq & \int\limits _{F.S.}
{dk_F(\theta )\over 2\pi} \left(
\int\limits _{-\Lambda }^{\Lambda}
{dk_{\perp } \over 2\pi }
c_{{\bf k}\sigma }
e^{ik_{\perp } {\bf n}(\theta )\cdot {\bf r}}
\right) e^{i {\bf k}_F(\theta )\cdot {\bf r}}
\nonumber \\
& {\equiv} & \int\limits _{F.S.}
{dk_F(\theta )\over 2\pi}
\psi _{\sigma , \theta } ({\bf r})
 e^{i {\bf k}_F(\theta )\cdot {\bf r}}.
\end{eqnarray}
\vskip 5 mm
\begin{figure}
\epsfig{file=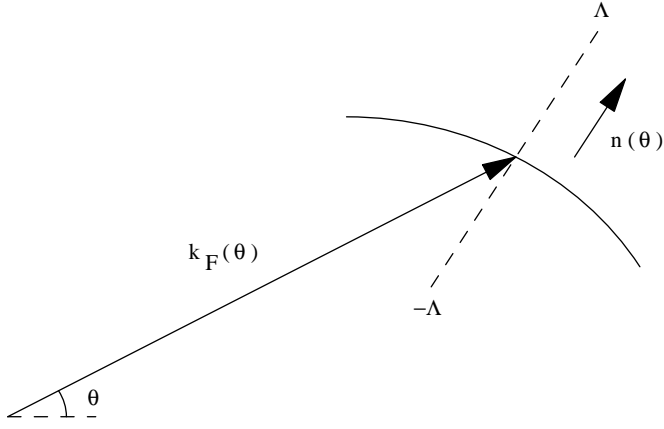,width=9.0cm}
\caption{Fermi-surface decomposition of the Fourier transform.}
\end{figure}
When we substitute this into (\ref{fullham}), we obtain
\begin{eqnarray}
\label{qcham}
H & = & \int d^2r \biggl[
\sum\limits_{\sigma }
\int\limits _{F.S.} {dk_F(\theta )\over 2\pi}
\psi ^{\dag }_{\sigma , \theta } ({\bf r})
{\bf v}_F (\theta ) \cdot (-i\nabla )
\psi _{\sigma , \theta } ({\bf r}) + \nonumber \\
& + & \int\limits _{F.S.}
{dk_F(\theta )\over 2\pi}
{dk_F(\theta ')\over 2\pi}
V(\theta , \theta ')
\psi ^{\dag }_{\uparrow \theta } ({\bf r})
\psi ^{\dag }_{\downarrow -\theta } ({\bf r})
\psi _{\downarrow -\theta '} ({\bf r})
\psi _{\uparrow \theta '} ({\bf r})\biggr], 
\end{eqnarray}
where ${\bf v}_F (\theta )$ is the Fermi velocity at the point
${\bf k}_F (\theta )$, and
$$
V(\theta , \theta ')\equiv \int d^2 r e^{-i({\bf k}_F(\theta ) -
{\bf k}_F(\theta '))\cdot {\bf r}} V({\bf r}).
$$
The derivation of (\ref{qcham}) is given in Appendix A. Note the linearized
kinetic energy in (\ref{qcham}), which will be crucial in the following.

The Hamiltonian (\ref{qcham}) gives rise to a partition function, which we can
write as a path integral over the fermion fields 
$\psi _{\sigma , \theta } ({\bf r})$. We can again decompose the interaction
by the HS transformation, {\it i.e.}, we can multiply the partition function
by the constant
\begin{eqnarray}
\label{HSconst}
\int {\cal D} \overline{\phi }_{\theta }({\bf r}) 
{\cal D} \phi _{\theta }({\bf r}) \exp \bigl[\int\limits_0^{\beta } d\tau
\int d^2r \int\limits_{F.S.} {dk_F(\theta )\over 2\pi }
{dk_F(\theta ')\over 2\pi } V(\theta ,\theta ') \times \nonumber \\
\times (\overline{\phi }_{\theta }({\bf r}) - 
\overline{\psi }_{\uparrow \theta } ({\bf r})
\overline{\psi }_{\downarrow -\theta } ({\bf r}))
(\phi _{\theta '}({\bf r})-
\psi _{\downarrow -\theta '}({\bf r}) \psi _{\uparrow \theta '} ({\bf r}))
\bigr].
\end{eqnarray}
In the mean-field approximation, the total free energy of the system equals
the free energy given by the (single-particle) Hamiltonian
\begin{eqnarray}
\label{mfham}
H & = & \int d^2r \biggl[
\int\limits _{F.S.} {dk_F(\theta )\over 2\pi}
\pmatrix{\psi ^{\dag }_{\uparrow \theta } ({\bf r}) &
\psi _{\downarrow -\theta } ({\bf r}) } 
\pmatrix{ {\bf v}_F (\theta ) \cdot (-i\nabla )
& \Delta _{\theta } ({\bf r}) \cr
\Delta ^* _{\theta } ({\bf r}) &
{\bf v}_F (\theta ) \cdot (i\nabla )}
\pmatrix{\psi _{\uparrow \theta } ({\bf r}) \cr
\psi ^{\dag }_{\downarrow -\theta } ({\bf r})}  - \nonumber \\
& - &  \int\limits _{F.S.}
{dk_F(\theta )\over 2\pi}
{dk_F(\theta ')\over 2\pi}
V(\theta , \theta ')
\phi ^* _{\theta } ({\bf r}) \phi _{\theta ' } ({\bf r}) \biggr],
\end{eqnarray}
upon minimization with respect to $\phi _{\theta } ({\bf r})$, where we
defined
\begin{equation}
\label{defdelta}
\Delta _{\theta } ({\bf r}) =
\int\limits _{F.S.} {dk_F(\theta )\over 2\pi}
V(\theta , \theta ') \phi _{\theta ' } ({\bf r}).
\end{equation}
We shall write explicit formulae for the total energy and free energy 
in Sections \ref{4} and \ref{5} (formulae (\ref{Efunc}) and (\ref{F})).
Here we just 
note that to calculate the single-particle contribution to the free energy,
we will have to find the spectra of the Andreev Hamiltonians labeled by
$\theta $, {\it i.e.}, we will need the energies $E_{\theta ,n}$ that
satisfy \cite{Andreev}
\begin{equation}
\label{andreev}
\pmatrix{ {\bf v}_F (\theta ) \cdot (-i\nabla )
& \Delta _{\theta } ({\bf r}) \cr
\Delta ^* _{\theta } ({\bf r}) &
{\bf v}_F (\theta ) \cdot (i\nabla )}
\pmatrix{f_{\theta ,n}({\bf r}) \cr g_{\theta ,n}({\bf r}) }
= E _{\theta ,n}
\pmatrix{f_{\theta ,n}({\bf r}) \cr g_{\theta ,n}({\bf r}) }.
\end{equation}
We note that the linear kinetic energy in (\ref{qcham}) makes 
this equation effectively one-dimensional, {\it i.e.}, 
an independent equation for each line in the direction ${\bf v}_F (\theta )$.
In the presence of the specularly reflecting boundary, we must find the
Andreev spectra along reflected lines such as the one in Fig. 1a. Equivalently,
we solve the equation on a straight line with the pairing potential 
$\Delta $ shown in Fig. 1b. This is intuitively obvious; a derivation is
given in Appendix B.

As we mentioned in the Introduction, the spectrum along each trajectory having
opposite signs of $\Delta $ at the two asymptotic ends will contain a
zero-bias state. Its wave function is, up to a normalization constant,
\begin{equation}
\label{bswavefn}
\pmatrix{f(\theta ,\rho ) \cr g (\theta ,\rho ) }_{ZBS}=
\pmatrix{1 \cr \mp i}
\exp (\mp \int\limits _0 ^{\rho } d \rho ' \Delta (\theta, \rho ')),
\end{equation}
where the upper (lower) sign corresponds to 
$\Delta (\theta ,\rho = -\infty ) < 0$, $\Delta (\theta ,\rho = +\infty )>0$
($\Delta (\theta ,\rho = -\infty )>0$, $\Delta (\theta ,\rho = +\infty )<0$),
so that the wave function is normalizable.
In our notation, we
 will freely interchange the dependence on ${\bf r}$ (actually only on $x$,
since the system is translationally invariant in the $y$-direction) with the
dependence on the angle $\theta $ and
the coordinate $\rho $ along the trajectory. Their relationship is obvious
from Fig. 1.

\subsection{Argument}
\label{argument}
We now have all the tools needed to demonstrate the $d+is$ transition in a
way that brings out its physical mechanism. We follow the same line of thought
as in Section \ref{2}: 
We go to the zero temperature, and look at the energy gains
and losses when the $s$-wave component of $\Delta $ appears.

For any $s$-wave pairing to appear, it is necessary that the part of
the functional integral (\ref{HSconst}) over the $s$-component of $\phi $
converge, {\it i.e.}, that $V$ on top of the dominant $d$-wave attraction 
contain
also an $s$-wave part
\footnote{We use again $|V_s|$ rather than $V_s$.},
$$
V(\theta ,\theta ') = V_d (\theta ,\theta ') -|V_s|.
$$
In this Subsection, we will show that this condition is also sufficient: At zero
temperature, the system will favor the $d+is$ state for an arbitrarily
weak attraction $V_s$.

With both $d$- and $s$-wave pairing present,
$$
\phi _{\theta }({\bf r}) = \phi _{d\theta }({\bf r}) + \phi _s({\bf r}).
$$
(The $s$-components of both $V$ and $\phi $ are angle-independent.) 
We begin with the second term in (\ref{mfham}), which then is
\begin{eqnarray}
&-& \int\limits _{F.S.} {dk_F(\theta )\over 2\pi}
{dk_F(\theta ')\over 2\pi}
(V_d(\theta ,\theta ')- |V_s|)
(\phi ^*_{d\theta }({\bf r}) + \phi ^*_s({\bf r}))
(\phi _{d\theta '}({\bf r}) + \phi _s({\bf r})) = \nonumber \\
&=& - \int\limits _{F.S.} {dk_F(\theta )\over 2\pi}
{dk_F(\theta ')\over 2\pi}
V_d(\theta ,\theta ') \phi ^*_{d\theta }({\bf r}) \phi _{d\theta '}({\bf r})
+ \nonumber \\
& + & |V_s| \int\limits _{F.S.} {dk_F(\theta )\over 2\pi}
{dk_F(\theta ')\over 2\pi} \phi ^*_s({\bf r}) \phi _s({\bf r}) ,
\end{eqnarray}
where we used the orthogonality of the $s$ and $d$ components:
$$
\int\limits _{F.S.} {dk_F(\theta )\over 2\pi}V_d(\theta ,\theta ')
= \int\limits _{F.S.} {dk_F(\theta ')\over 2\pi}V_d(\theta ,\theta ')
=\int\limits _{F.S.} {dk_F(\theta )\over 2\pi} \phi _{d\theta }({\bf r})=0.
$$
We can also split up (\ref{defdelta}) into components and define
\begin{eqnarray}
\Delta _{d\theta } ({\bf r}) & = &
\int\limits _{F.S.} {dk_F(\theta ')\over 2\pi}
V_d(\theta ,\theta ') \phi _{d\theta '}({\bf r}) \nonumber \\
\Delta _s ({\bf r}) & = &
\int\limits _{F.S.} {dk_F(\theta )\over 2\pi}
(-|V_s|)\phi _s({\bf r}) .
\end{eqnarray}
The $d$-component of $\Delta $ was established well above $T_{s}$, so
the change of the second term in (\ref{mfham}) due to the opening of
a (small) $s$-wave gap will be
\begin{equation}
\label{scondens}
{|\Delta _s({\bf r}) |^2 \over |V_s| }
\end{equation}
just as in the BCS case. Due to the translational invariance in
the $y$-direction, we will from now on write
$\Delta _s({\bf r}) \equiv \Delta _s (x)$. Along the quasiclassical
trajectory, $x$ depends on both $\rho $ and $\theta $ (see Fig.1),
so we will then write $\Delta _s(\theta ,\rho )$.

To examine the  effect of the small $s$ wave component on the quasiparticle
energies, we need to look at the change of the spectra of the $1D$ Andreev
problems
\begin{equation}
\label{dandreev}
\pmatrix{-i v_F (\theta ) \partial _{\rho } & \Delta _d(\theta, \rho ) \cr
\Delta _d (\theta, \rho ) & i v_F (\theta ) \partial _{\rho } }
\pmatrix{f_n (\theta, \rho ) \cr g_n (\theta, \rho ) } =
E _{\theta ,n} \pmatrix{f_n(\theta, \rho ) \cr g_n(\theta, \rho ) }.
\end{equation}
upon $\Delta _d(\theta, \rho ) \rightarrow \Delta _d(\theta, \rho )
+\Delta _s(\theta ,\rho ) $. As $\Delta _s$ is small, it can be treated as
a perturbation; then the change of the quasi-particle energies to the
lowest order is
\begin{eqnarray}
\label{deltaEn}
E^{(1)}_{\theta ,n} [\Delta _s]
& = & \int\limits _{-\infty }^{+\infty } d\rho
\pmatrix{f_n ^* (\theta, \rho ) & g_n^* (\theta, \rho )}
\pmatrix{0 & \Delta _s(\theta ,\rho ) \cr
\Delta ^*_s(\theta ,\rho ) & 0}
\pmatrix {f_n (\theta, \rho ) \cr g_n (\theta, \rho ) } = \nonumber \\
& = & \int\limits _{-\infty }^{+\infty } d\rho
[f_n ^* (\theta, \rho ) g_n (\theta, \rho ) \Delta _s(\theta ,\rho ) +
g_n^* (\theta, \rho ) f_n (\theta, \rho ) \Delta ^*_s(\theta ,\rho )].
\end{eqnarray}

Let us first look at the change of energy of the zero-energy bound states.
Then from (\ref{bswavefn})
\begin{equation}
\label{upperlower}
g_{ZBS}(\theta ,\rho )= \mp i f_{ZBS}(\theta ,\rho ),
\end{equation}
so
\begin{equation}
\label{Ebs}
E^{(1)}_{\theta ,ZBS}[\Delta _s] = \pm
\int\limits _{-\infty }^{+\infty } d\rho
|f(\theta, \rho ) |^2 2 Im \Delta _s(\theta ,\rho ),
\end{equation}
where the upper (lower) sign corresponds to the $+y$- ($-y$-)moving trajectory.
We notice several things by looking at (\ref{Ebs}):

\noindent
$\bullet $ It depends only on $Im \Delta _s$, since $Re \Delta _s$ just
changes the position of the node in the total $\Delta (\theta ,\rho )$, in
which case the bound state remains at zero energy. Hence, we will assume
$Re \Delta _s = 0$, and write $ \Delta _s(\theta ,\rho ) = is(\theta ,\rho )$.

\noindent
$\bullet $ $E^{(1)}_{\theta ,ZBS}[\Delta _s]$ is non-zero due to the form of the
bound-state wave function (\ref{upperlower}) and due to the fact that
$s(\theta ,\rho )$ does not change sign along the quasiclassical trajectory
(by virtue of the $s$-symmetry). Out of the two possibilities for the sign
of $s$, we will choose $s(\theta ,\rho ) > 0$ in the following, which means
all the $+y$-moving states are shifted up in energy, whereas the $-y$-moving
states are shifted down.

Since we are at zero temperature, only the states that move down from
zero energy will be occupied. We can then argue similarly as in the BCS
case: opening of the additional $s$-wave gap costs the system energy
$s^2 /|V_s|$ (from (\ref{scondens})) but the quasiparticles save energy
$\sim s$. The lowering of the quasiparticle energy is only linear in $s$,
$i.e.$, not as dramatic as the non-analytic decrease in the BCS case, but
nevertheless it beats the quadratic increase for small enough $s$. Thus,
for an arbitrarily small but non-zero interaction $|V_s|$, $s=0$ cannot be
a minimum of the total energy, and the additional $s$-wave gap phase shifted
by $\pi /2$ from the $d$-wave gap will appear. From the formula (\ref{Ebs}),
we see that the superconductor will benefit from opening up the gap only close
to the surface where $|f|^2$ is effectively non-zero, so the transition
into the $d+is$ state is a surface effect. The decay into the bulk will
be discussed more
quantitatively in the next section.

We should also note that the remaining states on the quasiclassical
trajectories do not change this situation, that is, they do not contribute
linearly to the change of the total quasiparticle energy. Due to the
time-reversal symmetry in the pure $d$-wave state, every state on a
given quasiclassical trajectory corresponds to a state of the same energy
on a reversed trajectory. Indeed, if we label the coordinate along the
trajectory reversed to the one in (\ref{dandreev}) as $\tilde \rho =-\rho $,
then the Hamiltonian on the reversed trajectory is
$$
\pmatrix{-i v_F (-\theta ) \partial _{\tilde \rho } &
\Delta _d(-\theta, \tilde \rho ) \cr
\Delta _d (-\theta, \tilde \rho ) &
i v_F (-\theta ) \partial _{\tilde \rho } } =
\pmatrix{i v_F (\theta ) \partial _{\rho } & \Delta _d(\theta, -\rho ) \cr
\Delta _d (\theta, -\rho ) & -i v_F (\theta ) \partial _{\rho } }
$$
since $v_F(-\theta )= v_F(\theta )$, so
$$
\pmatrix{f_n (-\theta, \tilde \rho ) \cr g_n (-\theta, \tilde \rho ) } =
\pmatrix{g_n (\theta, -\rho ) \cr f_n (\theta, -\rho ) }
$$
will also have energy $ E_{\theta ,n}$. Now
(\ref{deltaEn}) implies that to the first order,
a small $Im \Delta _s$ will shift the
energies of the two corresponding states by an equal amount with
opposite signs. Hence, the only way they can linearly contribute to the total
energy at $T=0$ is when one of them crosses zero and thus changes
its occupancy, which happens only when their original energy (in absolute value)
is smaller than the $s$-wave gap. But as $s\rightarrow 0$, there will
be fewer and fewer such states in smaller and smaller neighborhoods of the
$d$-wave nodes. It is only the ZBSs that change their occupancy
for arbitrarily small $s$. We thus conclude that the onset of the
transition into the $d+is$ state is driven by these states.

\section{$s$-wave gap at $T=0$}
\label{4}
In the study of the instability of the pure $d$ state in the last section,
we used first-order perturbation theory 
since $s\rightarrow 0$
at the onset of $d+is$. Now we will argue that this theory holds up to the the
actual value of $s$, $i.e.$ $s<<|\Delta _d|$. As we will show in 
Section \ref{5},
$s(T=0) \sim T_{s}$, the transition temperature into the $d+is$ state.
Because $T_{s}$ $\sim 7$ K, it is much smaller than $T_{d}$, the superconducting
transition temperature of the order of 100K, which sets the scale for
$\Delta _d$. Figure 3 shows the magnitude of the two gaps as a function
of angle around a quarter of the Fermi
surface. We see that the required inequality $s<<|\Delta _d|$ holds for most
of the Fermi surface except for small neighborhoods of the nodes. 
First-order perturbation theory certainly breaks down there, but upon
averaging over the Fermi surface, the nodes will only introduce an error of the
order $T_{s}/T_{d}$. Thus, we will use that theory to obtain $s(x)$ at $T=0$.
As discussed at the end of Subsection \ref{argument}, first-order perturbation
theory implies that we have to
look only at the zero-energy states. Also, since $s$ is a small perturbation,
we shall neglect its effect on $\Delta _d$.
\vskip 5 mm
\begin{figure}
\epsfig{file=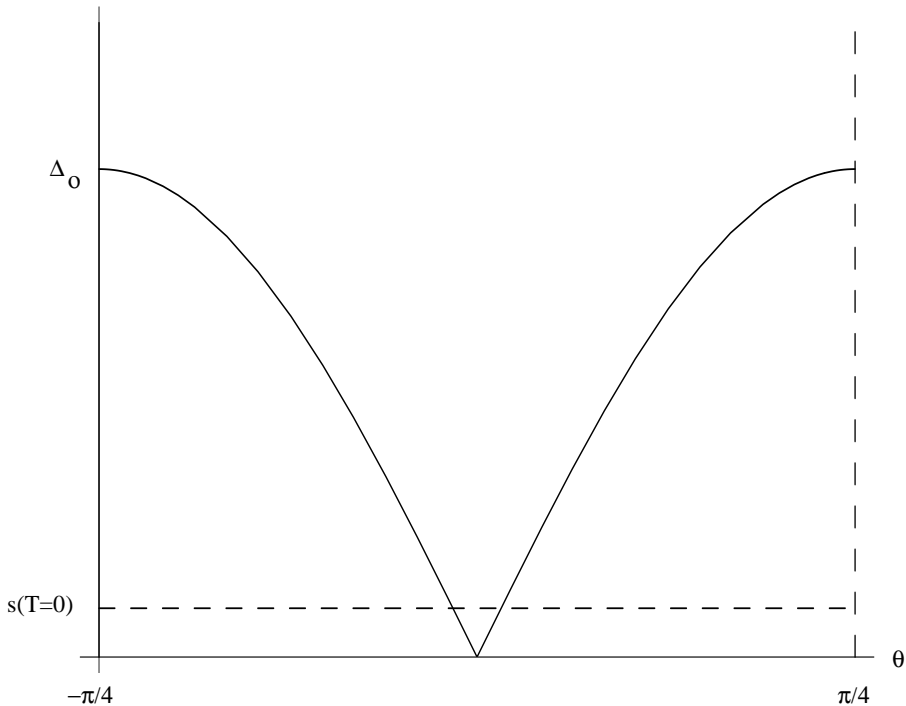,width=12.0cm}
\caption{The magnitude of the $d$- and $s$-wave order parameters around the
Fermi surface.}
\end{figure}
Now we can write down the energy due to $s$ per unit length of
the surface (the $y$-direction) as a functional of $s(x)$:
\begin{equation}
\label{Efunc}
E[s(x)]= \int\limits _0^{+\infty } dx {s^2(x) \over |V_s|}
+ \int\limits _{\theta \epsilon (-\pi /2,0)} {dk_F(\theta )\over 2\pi}
E_{\theta }[s(\theta ,\rho )]\cos \theta ,
\end{equation}
where $E_{\theta }[s]$ is given by (\ref{Ebs}); for the rest of this section,
we shall drop the superscript ``(1)'', since we shall be using only the
first-order formula. We freely interchange
$s(x)$ for $s(\theta ,\rho )$; the relation between the two is discussed
below (\ref{scondens}). Note the correct dimensions: the $x$-integration
makes $s^2/|V_s|$ from energy per unit area into energy per unit length.
In the second term, the integrand is energy and the dimension of the measure
is $k_F$, $i.e.$, inverse length. The extra factor of $\cos \theta$ in the
second integral accounts for the difference of the density of trajectories
along the $y$-direction compared to their angle-independent intrinsic density (measured
perpendicularly to their direction), as shown in Fig. 4.
In the second term in (\ref{Efunc}), we sum up only the occupied $-y$-moving
states for which $E_{\theta }<0$ according to (\ref{Ebs}).
\vskip 5mm
\begin{figure}
\epsfig{file=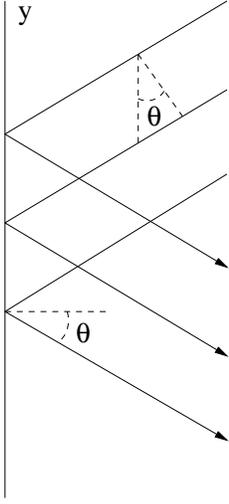,width=3.0cm}
\caption{The decrease of the density of the trajectories in the $y$-direction
by the factor $\cos \theta $.}
\end{figure}

We obtain $s(x)$ by minimizing (\ref{Efunc}). Let us first make an
order-of-magnitude estimate
\begin{equation}
\label{Eestim}
E[s]\sim \xi {s^2 \over |V_s|} - k_Fs,
\end{equation}
since $s$ will extend into the bulk only as far as the coherence length
$\xi = \hbar v_F /\Delta _o$ ($\Delta _o$ is the amplitude of the $d$ wave),
and from (\ref{Ebs}), we see $E_{\theta }[s] \sim s$. The angular averaging
will, up to numerical factors of order unity, multiply $E_{\theta }[s]$
by $k_F$. Minimization of (\ref{Eestim}) will give
\begin{equation}
\label{sestim}
s\sim {k_F |V_s| \over \xi}.
\end{equation}
By taking $s\sim 1$meV from the experiment, $k_F \sim 1$\AA $^{-1}$, and
$\xi \sim 10$\AA , we get an estimate for the strength of the $s$-wave
pairing
$$
|V_s| \sim 10 \mbox{meV\AA }^2.
$$

We minimize (\ref{Efunc}) exactly by solving
$$
{\delta E[s] \over \delta s(x)} = 0,
$$
$i.e.$,
$$
2{s(x) \over |V_s| }-
\int\limits _{\theta  \epsilon (-\pi /2,0)} {dk_F(\theta )\over 2\pi}
\cos \theta \int\limits _{-\infty }^{+\infty } d\rho
2|f(\theta ,\rho )|^2 {\delta s(\theta ,\rho )\over \delta s(x)} = 0.
$$
Now
\begin{eqnarray}
{\delta s(\theta ,\rho )\over \delta s(x)} & = &
\delta (x-\rho \cos \theta ) + \delta (x + \rho \cos \theta ) \nonumber \\
& = & {1\over \cos \theta }\left( \delta (\rho - {x\over \cos \theta })
+ \delta (\rho + {x\over \cos \theta }) \right),
\end{eqnarray}
since $\cos \theta > 0$, and $\rho $, unlike $x$, can be both positive and
negative. The factors of $\cos \theta $ cancel, and we obtain
\begin{equation}
\label{sT0}
s(x)=|V_s| \int\limits _{\theta  \epsilon (-\pi /2,0)} {dk_F(\theta )\over 2\pi}
\left( |f(\theta ,{x\over \cos \theta })|^2 +
|f(\theta ,-{x\over \cos \theta })|^2 \right).
\end{equation}
Physically, we get two terms on the right-hand side because for each angle
$\theta $, there are two trajectories contributing to $s$ at a given point
as shown in Fig. 5. 
\vskip 5 mm
\begin{figure}
\epsfig{file=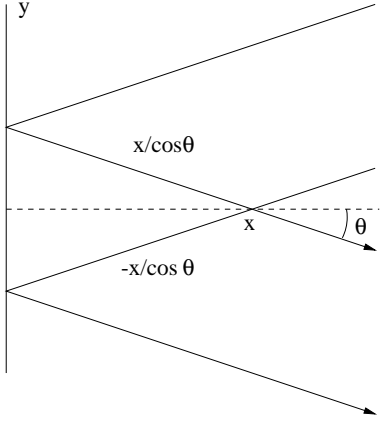,width=5.0cm}
\caption{Two contributions from the same $\theta $ to the pairing potential
at the point $x$.}
\end{figure}

We should remark here that we also obtain the formula (\ref{sT0}) when we
calculate the contribution from the occupied ($-y$-moving) bound states to
the pairing potential in the gap equation. This is done in Appendix C. The
result is
\begin{equation}
\label{sgapeqn}
\Delta _s (x)_{ZBS} = i|V_s|
\int\limits_{\theta \epsilon (-{\pi \over 2}, 0)}
{dk_F(\theta ) \over 2\pi } \bigl[
|f (\theta ,{x\over \cos \theta } )|^2 +
|f (\theta ,-{x\over \cos \theta } )|^2 \bigr]
\end{equation}
in agreement with (\ref{sT0}). This formula, however, shows more clearly
the internal consistency of the picture: For $\Delta _d$
in Fig. 1, the additional $is$ potential pushes down the $-y$-moving states
if $s>0$. As (\ref{sgapeqn}) shows, these states, in turn, give rise to
$\Delta _s = i\times positive$.

We should note here that $\Delta _s$ is absent on the right-hand side of
(\ref{sgapeqn}), so the gap equation in this case (unlike in the BCS
theory) is an explicit formula for the gap. The physical reason for this
is that $s(x)$ is considered small, so we neglect the change of the
bound-state wave functions due to its presence. The only effect of $s(x)$
we are taking into account is the change of the occupancy of the zero-energy
states, which, by (\ref{Ebs}), depends only on the sign of $s$, not on its
detailed shape. This is why $s(x)$ does not feed back into the right-hand
side of (\ref{sgapeqn}).

To estimate the decay of $\Delta _s$ into the bulk, we shall assume $\Delta _d$
to be constant in space and with the angular dependence
\begin{equation}
\Delta _{d,\theta }({\bf r}) = \Delta _o \sin 2\theta ,
\end{equation}
which should hold for
$$
x > \xi \equiv {\hbar v_F \over \Delta _o}.
$$
Also, we shall assume a spherical (circular) Fermi surface,
$$
dk_F(\theta ) = k_F d\theta .
$$
Then the wave function of the $-y$-moving bound states, 
including the normalization,
will be
\begin{equation}
\pmatrix{f(\theta ,\rho )\cr g(\theta ,\rho )} =
\sqrt{|\sin 2\theta |\over 2\xi } \pmatrix{1 \cr i}
e^{-|\rho \sin 2\theta |\over \xi },
\end{equation}
so
\begin{eqnarray}
s(x) & = & {k_F |V_s| \over \xi} \int\limits_{-\pi /2}^0
{d\theta \over 2\pi } 2\times {|\sin 2\theta |\over 2}
e^{-{2\over \xi} \left| {x\over \cos \theta } \sin 2 \theta \right|}
= \nonumber \\
& = & {k_F |V_s| \over \pi \xi } \int\limits_0^{\pi \over 2}
d\theta \sin \theta \cos \theta e^{-4 \sin \theta {x\over \xi}}.
\end{eqnarray}
We can do the integral by substitution $\sin \theta = t$, which
gives
\begin{equation}
\label{sexact}
s(x) ={k_F |V_s| \over \pi \xi } \left( -t {\xi \over 4x} -
\left( {\xi \over 4x} \right) ^2 \right) e^{-4t {x\over \xi}}
\Biggr] _ {t=0} ^1.
\end{equation}
We can neglect the contribution from the upper limit because it is effectively
non-zero only for $x<\xi /4$, where our assumption of constant $\Delta _d$ does
not hold. The lower limit should have been at $T_{s} /T_{d}$, rather than
at 0, to exclude the trajectories close to the nodes where the first-order
perturbation theory breaks down. That cuts off the lower-bound contribution
at $x \sim {T_{d} \over 4T_{s}} \xi \sim 100$\AA , beyond which we would
need a more refined theory for the behavior of the quasiparticles around the
nodes. 
For $x$ much smaller than this distance, we can neglect
the first term on the right hand side of (\ref{sexact}), and replace the
exponential by 1. We conclude, therefore, that
\begin{equation}
\label{sasymp}
s(x) \simeq {k_F |V_s| \xi \over 16 \pi x^2}
\end{equation}
for
$$
\xi < x << {T_{d}\over T_{s}} \xi.
$$
We see that $s=k_F |V_s| /\xi $ times a function that is of order unity
for $x<\xi$, and decays fast for $x>\xi$, as expected.

\section{Transition Temperature}
\label{5}
So far, we have shown the instability $d\rightarrow d+is$ only at
$T=0$. Just as in the BCS case, it remains to be demonstrated that the
transition temperature $T_{s}$ is finite. We therefore must study the
free energy of the system, which we obtain from (\ref{Efunc}) when we replace
$E_{\theta }[s]$ by ${\cal F} (E_{\theta }[s])$, the free energy of a single
fermion level (see (\ref{F1})), that is,
\begin{equation}
\label{F}
F[s] = \int\limits_0^{\infty } dx {s^2 (x) \over |V_s|} +
k_F \int\limits_{-\pi /2}^{\pi /2 } {d\theta \over 2\pi } \cos \theta
(-T) \ln (1+e^{-E_{\theta }[s]/T}).
\end{equation}
Minimization of this functional will give an equation for $s(x)$ that
again agrees with the contribution to the gap equation from the ZBSs.
As we see from (\ref{F}), the variational equation for $s$ will now be
very non-linear; it will no longer be an explicit formula for $s$. The
reason is that at finite temperatures, the occupancy of a given state 
depends on the value of its energy. Even in first-order perturbation
theory, this value depends on the shape of $s(x)$, not just its sign, so
$s(x)$ enters through the Fermi function into the right-hand side  of the
gap equation, making it non-linear and therefore difficult to solve.

We still can make an order-of-magnitude estimate of $F$ as follows
\begin{eqnarray*}
\ln (1+e^{-E_{\theta }[s]/T}) + \ln (1+e^{-E_{-\theta }[s]/T})
& = & \ln [(1+e^{-E_{\theta }[s]/T})(1+e^{E_{\theta }[s]/T})] \\
& = & \ln (2+2\cosh {E_{\theta }[s] \over T}) \\
& \simeq & \ln 4 + {1\over 4} \left( {E_{\theta }[s] \over T} \right) ^2
+ O (E_{\theta }[s] ^4) \\
& \sim & \ln 4 + {1\over 4} {s^2\over T^2} + O(s^4), 
\end{eqnarray*}
so
\begin{equation}
\label{festim}
F[s]-F[0] \sim s^2 \left( {\xi \over |V_s|} -{k_F \over T} \right)
+ O(s^4).
\end{equation}
From (\ref{festim}) we see that the system is unstable to the transition to the 
$d+is$ state below the temperature $T_{s} \sim k_F |V_s|/\xi $, which
is therefore of the same order of magnitude as $|\Delta _s|_{T=0}$.
(see (\ref{sestim})).

Following \cite{buchholtz}, we can trade the coupling constant $V_s$ for
the transition temperature, $T_{cs}$ of a BCS superconductor with this 
coupling,
$T_{cs}\sim e^{-1/|V_s|}.$ Then
\begin{equation}
T_s\sim {-1 \over \ln T_{cs}}.
\end{equation}
Hence, $T_s$ 
increases sharply close to $T_{cs}=0$, which is consistent with the 
numerical results \cite{fogelstrom,rainer}.

\section{Current}
\label{6}
To study the surface current in the $d+is$ state, we shall go back to
$T=0$ for simplicity. We observe that the states on the
$+y$-moving quasiclassical trajectory from Fig. 1 will,
upon the transition into the $d+is$ state with $s>0$, feel the pairing
potential shown in Fig. 6. As $\rho $ goes from $-\infty $ to $+\infty $,
the twist of the phase $\varphi $ of the order parameter is clockwise
(from $\pi $ to 0) for an $+y$-moving trajectory and counterclockwise
(from 0 to $\pi $) for a $-y$-moving one. In both cases, this implies
current flowing in the $-y$ direction. This agrees with our previous
calculations that showed that the $-y$- ($+y$-)moving bound states will be
(un)occupied if $s>0$.
\vskip 5 mm
\begin{figure}
\epsfig{file=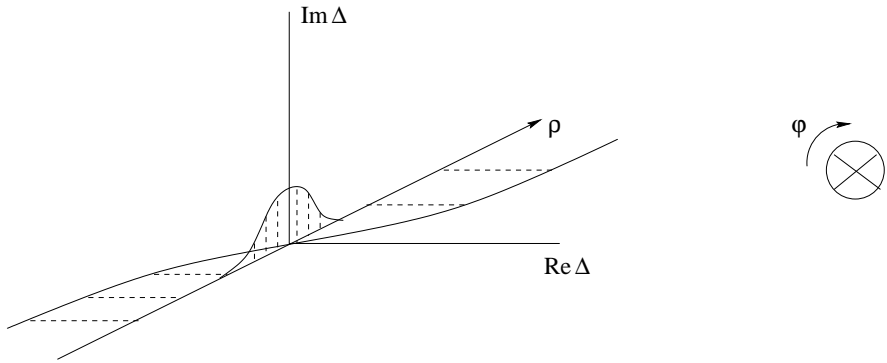,width=12.0cm}
\caption{The pairing potential along the trajectory in Fig. 1a). The 
corresponding twist of the phase of the order parameter is clockwise.}
\end{figure}

The agreement is quantitative, as we can easily check.
In the linearized Andreev formalism, the contribution to the current density
from a given state is
\begin{equation}
\label{j1dstates}
j^{(1d)}_n (\theta ,\rho )= ev_F
(|f_n(\theta ,\rho )|^2 + |g_n(\theta ,\rho )|^2 ),
\end{equation}
that is, the charge of the state times its (Fermi) velocity times the occupation
of that state. In our case, all of the current is carried by the occupied bound
states because the contributions from the remaining pairs of corresponding
countermoving states cancel each other out (see the end of Subsection
\ref{argument}).
To calculate the total
current density in the $y$-direction, we again have to include the 
contribution from both the incoming and the outgoing part of each 
$-y$-moving trajectory (see Fig. 5), 
and we have to project onto the $y$-direction
\begin{equation}
\label{jbs}
\left( j_{ZBS}(x) \right)_y
= k_F \int\limits_{-\pi /2}^0 {d\theta \over 2\pi} \sin \theta
\left[ j^{(1d)}_{ZBS}\left( \theta ,{x\over \cos \theta }\right) +
j^{(1d)}_{ZBS}\left( \theta ,-{x\over \cos \theta }\right) \right].
\end{equation}

 On the other hand, in terms of the one-dimensional density
$n^{(1d)}=k_F /\pi $ and the phase of the order
parameter $\varphi (\theta ,\rho )$ along the trajectory,
\begin{eqnarray}
\label{j1dop}
j^{(1d)} _{o.p.}(\theta ,\rho ) & = & e {1\over 2m} n^{(1d)} \partial _{\rho }
\varphi (\theta ,\rho ) =  \nonumber \\
& = & {ev_F \over 2\pi} \partial _{\rho } \varphi (\theta ,\rho ),
\end{eqnarray}
since for a spherical Fermi surface $v_F = k_F / m$.
The formula for the total surface-current 
density will be the same as (\ref{jbs}),
except that we now have to integrate over both $+y$- and $-y$-moving
trajectories
\begin{equation}
\label{jop}
\left( j_{o.p.}(x) \right)_y
= k_F \int\limits_{-\pi /2}^{\pi /2} {d\theta \over 2\pi} \sin \theta
\left[ j^{(1d)}_{o.p.}\left( \theta ,{x\over \cos \theta }\right) +
j^{(1d)}_{o.p.}\left( \theta ,-{x\over \cos \theta }\right) \right].
\end{equation}

We do not expect the current densities 
(\ref{jbs}) and (\ref{jop}) to be the same
at a given point because the formula (\ref{j1dop}) has corrections, which are
higher-order derivatives of $\varphi $. Those corrections will not,
however, contribute to the total surface current
\begin{equation}
I_y  =  \int_0^{\infty } dx j_y (x) ,
\end{equation}
which should then come out the same in the two calculations.
Indeed, the bound states give us
\begin{eqnarray}
(I_{ZBS})_y & = &  k_F \int\limits_{-\pi /2}^0 {d\theta \over 2\pi }
\sin \theta \cos \theta
\int\limits_{-\infty }^{+\infty } d\rho j^{(1d)}_{ZBS} (\theta ,\rho ),
\nonumber \\
& = & -{ev_Fk_F\over 4\pi},
\end{eqnarray}
since
$$
\int\limits_{-\infty }^{+\infty } d\rho j^{(1d)}_{ZBS} (\theta ,\rho )
=ev_F
$$
due to the normalization of the wave functions; the minus sign indicates
that the current is flowing in the $-y$ direction. 
The formula for the total current in
terms of the order-parameter phase is
\begin{equation}
(I_{o.p.})_y  =   k_F \int\limits_{-\pi /2}^{\pi /2} {d\theta \over 2\pi }
\sin \theta \cos \theta
\int\limits_{-\infty }^{+\infty } d\rho j^{(1d)}_{o.p.} (\theta ,\rho ).
\end{equation}
Now
\begin{eqnarray}
\label{intjop}
\int\limits_{-\infty }^{+\infty } d\rho j^{(1d)}_{o.p.} (\theta ,\rho )
& = & {ev_F\over \pi}
 (\varphi (\theta ,+\infty )- \varphi(\theta ,-\infty )) = \nonumber \\
& = & -{ev_F \over 2}\sgn (\theta ),
\end{eqnarray}
so
\begin{equation}
\label{agreement}
(I_{o.p.})_y = - {ev_Fk_F \over 4\pi} = (I_{ZBS})_y,
\end{equation}
since $\sgn (\theta ) \sin \theta $ is an even function, so the factor
1/2 in (\ref{intjop}) compensates for the doubling of the integration domain
of $\theta $ in (\ref{jop}) compared to (\ref{jbs}).
To get an order-of-magnitude estimate, we put
\begin{eqnarray*}
e& \sim & 10^{-19} \mbox{C} \\
v_F & \sim & 10 ^5 \mbox{m/s} \\
k_F & \sim & 10 ^{10} \mbox{m}^{-1}
\end{eqnarray*}
and get
$|I_y| \sim 10^{-5}$A  per CuO plane.
From the approximate form of the bound state wave functions introduced
in the previous section, we can also estimate the spatial distribution
of the current density
\begin{eqnarray}
j_y(x) & = & 4 ev_Fk_F \int\limits_{-\pi /2}^0 {d\theta \over 2\pi }
\sin \theta \left| f\left( \theta ,{x\over \cos \theta } \right) \right| ^2
= \nonumber \\
& = & -{4ev_Fk_F \over \xi } \int\limits_0^{\pi /2} {d\theta \over 2\pi }
\sin ^2 \theta \cos \theta e^{-4 {x\over \xi }\sin \theta } = \nonumber \\
& = & -{2ev_Fk_F \over \pi \xi } \int\limits_0^1 dt t^2 e^{-{4x\over \xi}t}.
\end{eqnarray}
By the same argument as presented in the last section, we find that for
$\xi < x << {T_{d}\over T_{s}} \xi ,$
\begin{equation}
\label{jasymp}
j_y(x) \simeq  -{ev_Fk_F \over 16\pi \xi } \left( {\xi \over x}\right) ^3.
\end{equation}
The extra power of $x$ in the denominator in (\ref{jasymp}), compared to
(\ref{sasymp}), comes from the directional sine in (\ref{jbs}).

The surface current induces magnetic field, which will be screened by the
diamagnetic current in the superconductor. The total current density therefore
is
\begin{equation}
(j_{tot}(x))_y = (j_{ZBS}(x))_y + (j_{dm}(x))_y.
\end{equation}
According to (\ref{jasymp}), the current is localized within the distance
$\sim \xi$ from the surface, which is much smaller than $\lambda $,
the in-plane penetration depth, because the cuprates are strongly type-2
superconductors. Hence, the diamagnetic response does not resolve the
internal structure of $(j_{ZBS}(x))_y$, and we can estimate
\begin{equation}
\label{jdm}
(j_{dm}(x))_y = (j_{dm}(0))_y e^{-x/\lambda }.
\end{equation}
\vskip 5 mm
\begin{figure}
\epsfig{file=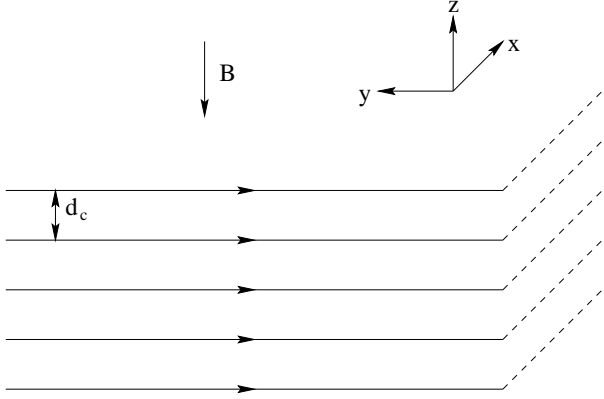,width=8.0cm}
\vskip 5mm
\caption{Side view of the $ab$ planes with the current flowing in the
$-y$ direction. The induced magnetic field is along the $c$-axis.}
\end{figure}
The surface current will be screened completely, because the magnetic
field it induces is smaller than $B_{c1}^c$, the lower critical field
in the $c$-direction. Indeed, even if the current flows in the same
direction along all the CuO planes (as shown in Fig. 7), the magnetic
field at distances $x>d_c$ (the interplane spacing $\sim 10$\AA ) from
the surface will be 
\begin{equation}
B={2\pi \over c}{I\over d_c} \sim 10^2 \mbox{G},
\end{equation}
which is smaller by an order of magnitude than $B_{c1}^c$ for YBCO
(see \cite{poole}). The complete screening implies
\begin{equation}
(I_{tot})_y \equiv \int\limits_0^{\infty} dx (j_{tot}(x))_y = 0,
\end{equation}
which together with (\ref{agreement}) gives
\begin{equation}
(j_{dm}(0))_y = {ev_Fk_F \over 4\pi \lambda }.
\end{equation}
For $\xi < x << {T_{d}\over T_{s}} \xi ,$ we can approximate the exponential
in (\ref{jdm}) by 1, so
\begin{equation}
(j_{tot}(x))_y \simeq {ev_Fk_F\over 16\pi }
\left[ {4\over \lambda}-{1\over \xi}\left( \xi \over x \right)^3 \right].
\end{equation}
This changes sign at distance
\begin{equation}
x_o \sim \xi \kappa ^{1/3},
\end{equation}
where $\kappa \equiv \lambda /\xi $ is the Ginzburg-Landau parameter. Due
to the one-third power, $x_o \sim \xi $ for reasonable values of $\kappa$
(say, between 50 and 500). This is consistent with the numerical
results \cite{fogelstrom}.

Note that the two calculations of the surface current agree
(see (\ref{agreement})) because the ZBSs moving in the direction of the 
current are shifted down in energy
and thus occupied, whereas those moving against the current are shifted up
and unoccupied. We wish to stress that this is exactly {\bf opposite}
 to the sign
of the Doppler shift: the states moving along the current would be
Doppler-shifted up, whereas those moving against the current would be
Doppler-shifted down.

This point is further supported by the analogy between the ZBSs and 
low-lying excitations in a core of an $s$-wave vortex. We consider an
idealized case: $\Delta =0$ inside the vortex (at distances from the
center smaller than $R$), 
and $|\Delta | = const $ outside with the phase winding counterclockwise
once around. We look at a quasiclassical trajectory passing close to the 
center of the vortex. We denote the coordinate along the trajectory as
$\rho $ again and the phase at the intersection point with the vortex edge
as $\varphi _{\pm }$, see Fig. 8. Then the energy of a low-lying excitation
moving from $\rho = -\infty $ to $\rho = +\infty $ on that trajectory is
\cite{vortex}
\begin{equation}
\label{Evortex}
E={v_F \over 4R} [(\varphi _+ - \varphi _-) - \pi ]_{mod \  2\pi}.
\end{equation}
For a trajectory passing through the center, 
$\varphi _+ - \varphi _- = \pi$, so $E=0$, and the low-lying excitation
is a ZBS. If we now shift the trajectory slightly to the left
as shown in Fig. 8, then $\varphi _+ - \varphi _- > \pi $ and $E>0$. In
a real vortex, $\Delta $ would be non-zero even inside, and the phase of
$\Delta $ would wind clockwise as we go from 
$\rho = -\infty $ to $\rho = +\infty $, so we are going against the current.
Moreover, $\varphi (\rho = +\infty )-\varphi (\rho =-\infty ) = \pi 
\mbox{ mod } 2\pi $, so $\Delta $ behaves the same way as in Fig. 6.
Hence, the $d\rightarrow d+is $ transition is analogous to shifting the
quasiclassical trajectory away from the vortex center. In both cases, the
ZBS will have a positive energy if it is moving against the current and
negative energy if it is moving in the direction of the current.
\vskip 5 mm
\begin{figure}
\epsfig{file=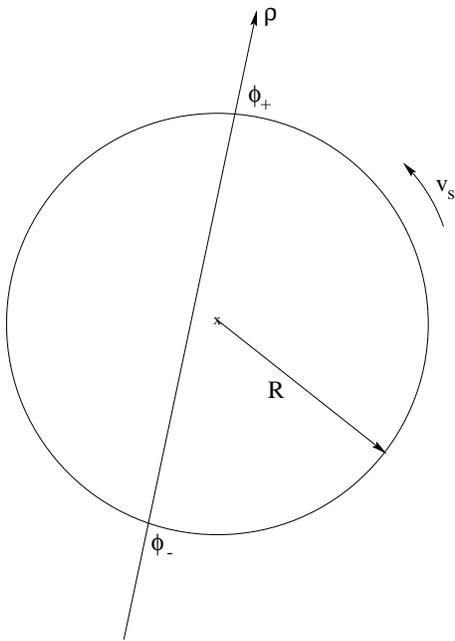,width=6.0cm}
\caption{A quasiclassical trajectory going through the core of an
$s$-wave vortex.}
\end{figure}

\section{Discussion}
\label{7}
In order to understand the basic physical mechanism of the 
$d\rightarrow d+is$ transition, we considered the change of the free energy
of a CuO half-plane when an $s$-wave component of $\Delta $ appears close
to the 110 surface. By the Hubbard-Stratonovich transformation and the
mean-field approximation, we decomposed the total free energy into the
contribution from the single-particle states, which is decreased by
Im $\Delta _s$, and the HS
term, which increases quadratically with $\Delta _s$. Using first-order
perturbation theory, we saw that the system favors the $d+is$ state
at $T=0$, and that the transition is driven by the zero-bias states. Based
on this argument and on the separation of energy scales associated with the
$d$- and $s$-wave components of $\Delta $, we then calculated $\Delta _s$ at
zero temperature and estimated the transition temperature. Finally, we 
discussed the surface current in the $d+is$ state. We saw that it is carried
by the occupied ZBSs; these states are {\bf not} Doppler-shifted by the
current.

\section*{Acknowledgements}
I want to give special thanks to M. Stone for suggesting the problem and
discussing it with me. 
I wish to thank A.J. Leggett for helping me figure out the diamagnetic
response in Sec.\ref{6}. I have also benefited from discussions with
I. Adagideli, H. Aubin, L.H. Greene, D.E. Pugel, 
R. Ramazashvili, S. Sachdev,
M. Turlakov,
and H. Westfahl. I am grateful to C. Elliott for proofreading the 
manuscript.
The project was supported by the grant NSF-DMR-98-17941.
\appendix
\section{Derivation of the Quasiclassical Hamiltonian}
\label{appA}
We derive (\ref{qcham}) from (\ref{fullham}) by expressing the original
field $\psi _{\sigma } ({\bf r})$ in terms of the slowly varying field
$\psi _{\sigma \theta } ({\bf r})$, see (\ref{qcfield}). We first substitute
into the kinetic-energy operator:
\begin{eqnarray}
\epsilon (-i \nabla ) \psi _{\sigma }({\bf r}) & = &
\int\limits _{F.S.} {dk_F(\theta )\over 2\pi}
e^{i {\bf k}_F(\theta )\cdot {\bf r}}
\epsilon ({\bf k}_F(\theta )-i \nabla )
\psi _{\sigma , \theta } ({\bf r}) = \nonumber \\
& = & \int\limits _{F.S.} {dk_F(\theta )\over 2\pi}
e^{i {\bf k}_F(\theta )\cdot {\bf r}} 
 \left[ \epsilon ({\bf k}_F(\theta )) +
\nabla _{\bf k} \epsilon ({\bf k})|_{{\bf k}_F} (\theta ) \cdot
(-i\nabla ) + \cdots \right]
\psi _{\sigma , \theta } ({\bf r}) \simeq \nonumber \\
&\simeq & \int\limits _{F.S.} {dk_F(\theta )\over 2\pi}
e^{i {\bf k}_F(\theta )\cdot {\bf r}}
{\bf v}_F (\theta ) \cdot (-i\nabla )
\psi _{\sigma , \theta } ({\bf r}),
\end{eqnarray}
since
$$
\epsilon ({\bf k}_F) = 0 \mbox{, and}
\nabla _{\bf k} \epsilon ({\bf k})|_{{\bf k}_F} = {\bf v}_F.
$$
The kinetic energy therefore is
\begin{eqnarray}
\int d^2 r \sum\limits_{\sigma = \uparrow , \downarrow}
\psi ^{\dag} _{\sigma } ({\bf r})
\epsilon (-i \nabla ) \psi _{\sigma }({\bf r})& = &
 \int d^2 r \sum\limits_{\sigma }
\int\limits _{F.S.} {dk_F(\theta )\over 2\pi}
\int\limits _{F.S.} {dk_F(\theta' )\over 2\pi}
e^{i({\bf k}_F(\theta )-{\bf k}_F(\theta ' ))\cdot {\bf r}}
\times \nonumber \\
& \times &
\psi ^{\dag }_{\sigma , \theta '} ({\bf r})
{\bf v}_F (\theta ) \cdot (-i\nabla )
\psi _{\sigma , \theta } ({\bf r}).
\end{eqnarray}
Now $e^{i({\bf k}_F(\theta )-{\bf k}_F(\theta ' ))\cdot {\bf r}}$
oscillates with a wavelength much shorter than the lengthscale
on which $\psi _{\sigma , \theta } ({\bf r})$ changes.
Thus, the integral will be zero unless
${\bf k}_F(\theta )-{\bf k}_F(\theta ' )=0$, so we can
effectively drop one integration over the Fermi surface, and
obtain the kinetic energy of the form
\begin{equation}
\int d^2 r \sum\limits_{\sigma }
\int\limits _{F.S.} {dk_F(\theta )\over 2\pi}
\psi ^{\dag }_{\sigma , \theta } ({\bf r})
{\bf v}_F (\theta ) \cdot (-i\nabla )
\psi _{\sigma , \theta } ({\bf r}).
\end{equation}

The potential energy is now given by
\begin{eqnarray}
& & \int d^2 rd^2r' {dk_F(\theta _1)\over 2\pi}
{dk_F(\theta _2)\over 2\pi}
{dk_F(\theta _3)\over 2\pi}
{dk_F(\theta _4)\over 2\pi}
V({\bf r-r'}) \times \nonumber \\
& & \quad \times
\exp[i(-{\bf k}_F(\theta _1) \cdot {\bf r}
-{\bf k}_F(\theta _2) \cdot {\bf r'}
+{\bf k}_F(\theta _3) \cdot {\bf r'}
+{\bf k}_F(\theta _4) \cdot {\bf r})] \times
\psi ^{\dag }_{\uparrow \theta _1} ({\bf r})
\psi ^{\dag }_{\downarrow \theta _2} ({\bf r'})
\psi _{\downarrow \theta _3} ({\bf r'})
\psi _{\uparrow \theta _4} ({\bf r}) 
\nonumber \\[.25in]
& & \approx
\int d^2 rd^2r' {dk_F(\theta _1)\over 2\pi}
{dk_F(\theta _2)\over 2\pi}
{dk_F(\theta _3)\over 2\pi}
{dk_F(\theta _4)\over 2\pi}
V({\bf r-r'})
e^{i({\bf k}_F(\theta _2) -
{\bf k}_F(\theta _3))\cdot ({\bf r-r'})}
\times \nonumber \\
& & \quad \times
\psi ^{\dag }_{\uparrow \theta _1} ({\bf r})
\psi ^{\dag }_{\downarrow \theta _2} ({\bf r})
\psi _{\downarrow \theta _3} ({\bf r})
\psi _{\uparrow \theta _4} ({\bf r})
\times
\exp[i(-{\bf k}_F(\theta _1)
-{\bf k}_F(\theta _2)
+{\bf k}_F(\theta _3)
+{\bf k}_F(\theta _4)) \cdot {\bf r}]
 \nonumber \\[.25in]
& & =\int d^2 rd^2r' {dk_F(\theta _1)\over 2\pi}
{dk_F(\theta _2)\over 2\pi}
{dk_F(\theta _3)\over 2\pi}
{dk_F(\theta _4)\over 2\pi}
V(\theta _2, \theta _3) \times \nonumber \\
& & \quad \times
\psi ^{\dag }_{\uparrow \theta _1} ({\bf r})
\psi ^{\dag }_{\downarrow \theta _2} ({\bf r})
\psi _{\downarrow \theta _3} ({\bf r})
\psi _{\uparrow \theta _4} ({\bf r})
\times 
\exp[i(-{\bf k}_F(\theta _1)
-{\bf k}_F(\theta _2)
+{\bf k}_F(\theta _3)
+{\bf k}_F(\theta _4)) \cdot {\bf r}],
\end{eqnarray}
where we used the assumption that $V$ changes on a much shorter length scale
than $\psi _{\sigma , \theta } ({\bf r})$, performed the $r'$-integration,
and introduced the Fourier transform
$$
V(\theta , \theta ')\equiv \int d^2 r e^{-i({\bf k}_F(\theta ) -
{\bf k}_F(\theta '))\cdot {\bf r}} V({\bf r}).
$$
Since $V({\bf r})=V(-{\bf r})$, we see that
$V(\theta , \theta ') = V(\theta ', \theta )$. Again, the integral vanishes
unless the sum of the four momenta is zero. Out of the various ways that
this may happen, we pick only the one that contributes to the singlet pairing,
namely
\begin{eqnarray}
{\bf k}_F(\theta _1) + {\bf k}_F(\theta _2) & = &
{\bf k}_F(\theta _3)
+{\bf k}_F(\theta _4) = 0 , i.e. \nonumber \\
\theta _1 + \theta _2 & = & \theta _3 + \theta _4
=0 \ \mbox{mod} \ 2\pi.
\end{eqnarray}
Since we now have two constraints, we can drop two Fermi-surface integrations,
and obtain the potential energy of the form
$$
\int d^2 r {dk_F(\theta )\over 2\pi}
{dk_F(\theta ')\over 2\pi}
V(\theta , \theta ')
\psi ^{\dag }_{\uparrow \theta } ({\bf r})
\psi ^{\dag }_{\downarrow -\theta } ({\bf r})
\psi _{\downarrow -\theta '} ({\bf r})
\psi _{\uparrow \theta '} ({\bf r}).
$$

The total Hamiltonian in the quasiclassical approximation then is
\begin{eqnarray}
\label{qchamApp}
H & = & \int d^2r \biggl[
\sum\limits_{\sigma }
\int\limits _{F.S.} {dk_F(\theta )\over 2\pi}
\psi ^{\dag }_{\sigma , \theta } ({\bf r})
{\bf v}_F (\theta ) \cdot (-i\nabla )
\psi _{\sigma , \theta } ({\bf r}) + \nonumber \\
& + & \int\limits _{F.S.}
{dk_F(\theta )\over 2\pi}
{dk_F(\theta ')\over 2\pi}
V(\theta , \theta ')
\psi ^{\dag }_{\uparrow \theta } ({\bf r})
\psi ^{\dag }_{\downarrow -\theta } ({\bf r})
\psi _{\downarrow -\theta '} ({\bf r})
\psi _{\uparrow \theta '} ({\bf r})\biggr] .
\end{eqnarray}
The derivation of (\ref{qchamApp}) from (\ref{fullham}) is far from rigorous.
We could give a somewhat better, although much longer, argument. We believe
the approximations used here are equivalent to the approximation in the
Eilenberger formalism, because the Eilenberger equations can now be
rigorously (apart from the mean-field approximation) derived from
(\ref{qchamApp}).

\section{Boundary Conditions at the Surface}
\label{appB}
We show the effect of the boundary on the Andreev spectrum. In general,
the boundary will cause mixing of different $\theta $'s. For each
$\theta $, though, we have a different Andreev equation, so adding the
solutions of (\ref{andreev}) for different $\theta $'s does not make
sense. However, the Andreev wave functions describe only the slow variation
of our excitations (changes on the length-scale $\xi $). The full wave
functions containing the rapid oscillations as well are
\begin{equation}
\label{fullsoln}
\pmatrix{f_{\theta ,n}({\bf r}) \cr g_{\theta ,n}({\bf r}) }
e^{i{\bf k}_F(\theta )\cdot {\bf r}},
\end{equation}
and these describe the single-particle excitations of the {\it same}
Hamiltonian (\ref{fullham}) (in the mean-field approximation), so those
can be added. If we assume a specularly reflecting
boundary, then the wave function will contain only two terms:
$$
\pmatrix{f_{\theta _{in},n}({\bf r}) \cr g_{\theta _{in},n}({\bf r}) }
e^{i{\bf k}_F({\theta _{in}})\cdot {\bf r}} +
\pmatrix{f_{\theta _{out},n}({\bf r}) \cr g_{\theta _{out},n}({\bf r}) }
e^{i{\bf k}_F({\theta _{out}})\cdot {\bf r}},
$$
such that
\begin{equation}
\label{inout}
\theta _{in} + \theta _{out} = \pi \ \mbox{mod} \ 2\pi,
\end{equation}
since the angles are measured from the positive-$x$ semiaxis,
see Fig. 1a. The Dirichlet boundary condition gives
$$
\pmatrix{f_{\theta _{out},n}({\bf r}) \cr g_{\theta _{out},n}({\bf r}) } =
\pmatrix{f_{\theta _{in},n}({\bf r}) \cr g_{\theta _{in},n}({\bf r}) }
\left(-e^{i({\bf k}_F({\theta _{in}})-{\bf k}_F({\theta _{out}}))
\cdot {\bf r}} \right) _{{\bf r}\epsilon surface},
$$
whereas the Neumann boundary condition gives
$$
\pmatrix{f_{\theta _{out},n}({\bf r}) \cr g_{\theta _{out},n}({\bf r}) } =
\pmatrix{f_{\theta _{in},n}({\bf r}) \cr g_{\theta _{in},n}({\bf r}) }
\left(e^{i({\bf k}_F({\theta _{in}})-{\bf k}_F({\theta _{out}}))
\cdot {\bf r}} \right) _{{\bf r}\epsilon surface}
$$
since
$$
{\bf n}\cdot ({\bf k}_F({\theta _{in}})-{\bf k}_F({\theta _{out}})) = 0
$$
for ${\bf n}$ perpendicular to the surface, and we used
$$
k_F \pmatrix{f \cr g}>> {\bf n}\cdot \nabla _r \pmatrix{f \cr g}
$$
to neglect the gradient of the Andreev wave function.
Since (\ref{andreev}) is linear, we can drop multiplicative constants
and simply assume
\begin{equation}
\label{matching}
\pmatrix{f_{\theta _{out},n}({\bf r}) \cr g_{\theta _{out},n}({\bf r}) } =
\pmatrix{f_{\theta _{in},n}({\bf r}) \cr g_{\theta _{in},n}({\bf r}) }
\end{equation}
at the surface for either choice of the boundary condition.
As $\theta _{in}$ is uniquely determined by $\theta _{out}$ through the
relation (\ref{inout}), we shall label the potential $\Delta $ along
the trajectory as well as the solutions of the corresponding Andreev
equation by $\theta _{out}$. We shall drop the subscript ``$out$'' everywhere
except in Appendix C, where we will need to distinguish $\theta _{out}$,
the label for a trajectory as in Fig. 1, from $\theta $, the label for a 
position on the Fermi surface as in Fig. 2.

\section{Gap Equation}
We obtain the gap equation by substituting for $\phi _{\theta }({\bf r})$
in (\ref{defdelta}) its mean-field value, that is, the pairing amplitude
$\phi _{\theta }({\bf r}) \equiv
\langle \psi _{\downarrow -\theta} ({\bf r})
\psi _{\uparrow \theta} ({\bf r}) \rangle$. To calculate this amplitude,
we expand the field operators into energy eigenstates
\begin{equation}
\label{expand}
\pmatrix{ \psi _{\uparrow \theta} ({\bf r}) \cr
\psi ^{\dag } _{\downarrow -\theta} ({\bf r}) }
= \sum\limits _n \gamma _{\theta ,n}
\pmatrix{f_n(\theta ,\rho ) \cr g_n(\theta ,\rho )}.
\end{equation}
Equation (\ref{expand}) gives at $T=0$
\begin{eqnarray}
\langle \psi _{\downarrow -\theta} ({\bf r})
\psi _{\uparrow \theta} ({\bf r}) \rangle
& = &
\sum\limits_{n, n'} f_n(\theta ,\rho ) g^* _{n'} (\theta ,\rho )
\langle \gamma ^{\dag } _{\theta ,n'} \gamma _{\theta ,n} \rangle
= \nonumber \\
& = &
\sum\limits _n \Theta (-E_{\theta ,n})
f_n(\theta ,\rho ) g^* _n (\theta ,\rho ).
\end{eqnarray}
Note that in (\ref{expand}), we explicitly sum over both positive and
negative energies, unlike the Bogoliubov-de Gennes (BdG) 
formalism where we can sum
over positive energies only, using the fact that
\begin{equation}
\label{symmetry}
\pmatrix{u_n ({\bf r}) \cr v_n ({\bf r})} \ \mbox{and}
\pmatrix{-v^*_n({\bf r}) \cr u^*_n({\bf r})}
\end{equation}
are both solutions of the BdG equations with energies equal in absolute value
and opposite in sign. However, this symmetry is lost here because the Andreev
wave functions corresponding to the BdG wave functions (\ref{symmetry}) live
on different quasiclassical trajectories.

Close to the surface, we have to remember again that each line contributes to
the pairing amplitude at a given point for two directions $\theta $; see
Fig. 5. We will, therefore, have to distinguish between the label of the
trajectory $\theta _{out}$ and the label for the pairing amplitude $\theta $.
Specifically,
\begin{eqnarray}
\label{21corresp}
\theta _{out} & = & \theta \qquad \mbox{for} \ \theta \epsilon (-{\pi \over 2}, 0)
\ \mbox{and} \nonumber \\
\theta _{out} & = & -\pi - \theta  \qquad \mbox{for} \
\theta \epsilon (-\pi,-{\pi \over 2}),
\end{eqnarray}
so the contribution to the pairing amplitude from the $-y$-moving bound states
will be
\begin{eqnarray}
\label{pairingout}
\langle \psi _{\downarrow -\theta} ({\bf r})
\psi _{\uparrow \theta} ({\bf r}) \rangle _{ZBS}
& = &
f (\theta _{out},{x\over \cos \theta _{out}} )
g^* (\theta _{out},{x\over \cos \theta _{out}} ) = \nonumber \\
& = &
f (\theta ,{x\over \cos \theta } )
g^* (\theta ,{x\over \cos \theta } ) \
\mbox{for} \ \theta \epsilon (-{\pi \over 2}, 0)
\end{eqnarray}
and
\begin{eqnarray}
\label{pairingin}
\langle \psi _{\downarrow -\theta} ({\bf r})
\psi _{\uparrow \theta} ({\bf r}) \rangle _{ZBS}
& = &
f (\theta _{out},-{x\over \cos \theta _{out}} )
g^* (\theta _{out},-{x\over \cos \theta _{out}} ) = \nonumber \\
& = &
f (-\pi -\theta ,{x\over \cos \theta } )
g^* (-\pi -\theta ,{x\over \cos \theta } ) \
\mbox{for} \ \theta \epsilon (-\pi ,-{\pi \over 2}).
\end{eqnarray}
Substituting (\ref{pairingout}) and (\ref{pairingin}) into 
(\ref{defdelta}) gives
\begin{eqnarray}
\Delta _{\theta } (x)_{ZBS} & = &
\int\limits _{\theta ' \epsilon (-\pi ,0)}
{dk_F(\theta ') \over 2\pi } V(\theta ,\theta ')
\langle \psi _{\downarrow -\theta '} ({\bf r})
\psi _{\uparrow \theta '} ({\bf r}) \rangle _{ZBS} = \nonumber \\
& = & \int\limits_{\theta '_{out}\epsilon (-{\pi \over 2}, 0)}
{dk_F(\theta '_{out}) \over 2\pi } \bigl[
V(\theta ,\theta '_{out}) (-i)
|f (\theta '_{out},{x\over \cos \theta '_{out}} )|^2 + \nonumber \\
& + & V(\theta ,-\pi -\theta '_{out}) (-i)
|f (\theta '_{out},-{x\over \cos \theta '_{out}} )|^2 \bigr],
\end{eqnarray}
where
$$
V(\theta ,\theta '_{out}) = -|V_s| + V_d(\theta ,\theta '_{out}),
$$
and we used (\ref{upperlower}). The contribution to the $s$-wave
component of the pairing potential from the occupied bound states therefore
is
\begin{equation}
\Delta _s (x)_{ZBS} = i|V_s|
\int\limits_{\theta '_{out}\epsilon (-{\pi \over 2}, 0)}
{dk_F(\theta '_{out}) \over 2\pi } \bigl[
|f (\theta '_{out},{x\over \cos \theta '_{out}} )|^2 +
|f (\theta '_{out},-{x\over \cos \theta '_{out}} )|^2 \bigr]
\end{equation}

For the calculation of the $d$-wave component of $\Delta _{ZBS}$, we will
assume that the unperturbed $\Delta _d$ is antisymmetric around its vertical
node,
\begin{equation}
\label{antisym}
\Delta _{d,\theta}({\bf r}) = -\Delta _{d,-\pi -\theta } ({\bf r}).
\end{equation}
Presumably, $\Delta _d$ arises from an antisymmetric interaction
\begin{equation}
V_d(\theta ,\theta ') = -V_d(\theta ,-\pi -\theta ').
\end{equation}
Along the quasiclassical trajectory, (\ref{antisym}) means
\begin{equation}
\Delta _d (\theta ,\rho )= -\Delta _d (\theta ,-\rho ),
\end{equation}
which, by (\ref{bswavefn}), implies
\begin{equation}
|f(\theta ,\rho )|^2 = |f(\theta ,-\rho )|^2.
\end{equation}
Thus, under these assumptions,
\begin{eqnarray}
\Delta _{d,\theta }(x) _{ZBS} & = &
(-i)\int\limits_{\theta '_{out}\epsilon (-{\pi \over 2}, 0)}
{dk_F(\theta '_{out}) \over 2\pi } \bigl[
V_d(\theta ,\theta '_{out}) + V_d(\theta ,-\pi -\theta '_{out}) \bigr]
|f (\theta '_{out},{x\over \cos \theta '_{out}} )|^2 = \nonumber \\
& = & 0.
\end{eqnarray}

\end{document}